\documentclass[aps,prl,amsfonts,twoside,twocolumn,amssymb,superscriptaddress]{revtex4}

\usepackage[latin1]{inputenc}
\usepackage[english]{babel}
\usepackage[draft]{hyperref}
\usepackage{graphicx}
\usepackage{amsmath}
\usepackage{color}
\usepackage{float}
\usepackage{units}
\graphicspath{{images/},{figures/}}
\usepackage{amsmath, amsthm, amssymb}

\newcommand{\eref}[1]{Eq.~(\ref{#1})} 
\newcommand{\fref}[1]{Fig.~\ref{#1}}  

\newcommand{\rmi}{\mathrm{i}}
\newcommand{\abs}[1]{\Big|#1\Big|}

\newcommand{\rbs}{r_{\rm{bs}}}
\newcommand{\tbs}{t_{\rm{bs}}}

\newcommand{\rmem}{r_{\rm{m}}}
\newcommand{\tmem}{t_{\rm{m}}}

\begin{document}

\title{Interferometer readout-noise below the Standard Quantum Limit of a membrane}

\author{T. Westphal}
\affiliation{Institut f\"{u}r Gravitationsphysik, Leibniz Universit\"{a}t Hannover and \\ Max-Planck-Institut f\"{u}r Gravitationsphysik (Albert-Einstein-Institut), 30167 Hannover, Germany}

\author{D. Friedrich}
\affiliation{Institut f\"{u}r Gravitationsphysik, Leibniz Universit\"{a}t Hannover and \\ Max-Planck-Institut f\"{u}r Gravitationsphysik (Albert-Einstein-Institut), 30167 Hannover, Germany}

\author{H. Kaufer}
\affiliation{Institut f\"{u}r Gravitationsphysik, Leibniz Universit\"{a}t Hannover and \\ Max-Planck-Institut f\"{u}r Gravitationsphysik (Albert-Einstein-Institut), 30167 Hannover, Germany}

\author{K. Yamamoto}
\affiliation{Institute for Cosmic Ray Research, The University of Tokyo, 5-1-5 Kashiwa-no-Ha, Kashiwa, Chiba 277-8582, Japan}

\author{S.~Go{\ss}ler}
\affiliation{Institut f\"{u}r Gravitationsphysik, Leibniz Universit\"{a}t Hannover and \\ Max-Planck-Institut f\"{u}r Gravitationsphysik (Albert-Einstein-Institut), 30167 Hannover, Germany}

\author{H.~M\"uller-Ebhardt}
\affiliation{Institut f\"{u}r Gravitationsphysik, Leibniz Universit\"{a}t Hannover and \\ Max-Planck-Institut f\"{u}r Gravitationsphysik (Albert-Einstein-Institut), 30167 Hannover, Germany}

\author{S.~L.~Danilishin}
\affiliation{Department of Physics, Moscow State University, Moscow RU-119992, Russia}

\author{F.~Ya.~Khalili}
\affiliation{Department of Physics, Moscow State University, Moscow RU-119992, Russia}

\author{K.\ Danzmann}
\affiliation{Institut f\"{u}r Gravitationsphysik, Leibniz Universit\"{a}t Hannover and \\ Max-Planck-Institut f\"{u}r Gravitationsphysik (Albert-Einstein-Institut), 30167 Hannover, Germany}

\author{R.\ Schnabel}
\affiliation{Institut f\"{u}r Gravitationsphysik, Leibniz Universit\"{a}t Hannover and \\ Max-Planck-Institut f\"{u}r Gravitationsphysik (Albert-Einstein-Institut), 30167 Hannover, Germany}


\begin{abstract}
Here we report on the realization of a Michelson-Sagnac interferometer whose purpose is the precise characterization of the motion of 
membranes showing significant light transmission. Our interferometer has a readout noise spectral density (imprecision) of 
 $3 \times 10^{-16} \rm{m}/\!\sqrt{\rm{Hz}}$ at frequencies around the fundamental resonance of  a SiN$_{\rm{x}}$ membrane at about 100\,kHz, without using optical cavities. The readout noise demonstrated is more than 16\,dB below {the peak value of the} membrane's standard quantum limit (SQL). This reduction is significantly higher than those of previous works with nano-wires 
[Teufel \emph{et al.}, Nature Nano. {\bf 4}, 820 (2009);  Anetsberger \emph{et al.}, Nature Phys. {\bf 5}, 909 (2009)]. 
We discuss the meaning of the SQL for force measurements and its relation to the readout performance and conclude that neither our nor previous experiments achieved a \emph{total} noise spectral density as low as the SQL. 

\end{abstract}

\maketitle

Systems composed of mechanical probes and laser light 
are able to perform ultra-sensitive measurements of external forces by monitoring the probes' displacements. A prominent example is a gravitational wave detector, which uses laser light and quasi free-falling test mass mirrors to search for changes in the far-field of accelerated gravitational sources \cite{Abramovici92}. The standard quantum limit (SQL) is a consequence of Heisenberg's Uncertainty Principle \cite{Heisenberg} and sets a limit to the sensitivity of a continuous measurement (monitoring) of 
a quantity that does not commute with itself at different times \cite{Bra1968}. 
With respect to the internal degrees of freedom of opto-mechanical measurement devices, the SQL also defines a benchmark noise spectral density at which the door is opened to new quantum experiments such as the generation of opto-mechanical entanglement \cite{VGFBTGVZA07} and also purely mechanical entanglement \cite{Helge08}.
  
So far,  
{a total noise spectral density as low as the SQL of a force measurement has not yet been achieved, since the thermal excitation of the mechanical probe is usually far above the SQL. 
}
In general, reaching the SQL requires 
(i) the acceleration noise of the mechanical probe being dominated by quantum back-action due to the (optical) readout,
(ii) the readout noise being dominated by (photon) shot-noise, and
(iii) the readout power being optimized such that for uncorrelated quantum back-action noise and quantum readout noise the two are of identical size, and thus provide one half of the SQL's noise power each \cite{Braginsky95}. 
In case of quantum correlations between the readout quadrature amplitude and its orthogonal quadrature, condition (iii) may vary, even going beyond the SQL is possible \cite{KLMTV01}. 
{Furthermore, it is important to note that, generally, the SQL does not have a white spectrum but depends on the observation frequency. It also depends on the probe's dynamics and on the observable chosen.  In particular, in the case of a harmonic oscillator, the SQL of a position measurement shows a peak at the mechanical resonance, while the SQL for a force measurement has a minimum here as seen in equations~(\ref{SQLx}) and (\ref{SQLF}). }

Recently, a microwave \cite{Teu09} and an optical readout imprecision \cite{Kipp2009} was demonstrated below the SQL's \emph{peak} values of nano-wires. The SQL of the measurement of an external force itself was, however, not reached, because the probe's motion was dominated by thermal excitation, and the imprecision was not photon shot-noise limited. Although neither experiment reached the SQL, the number of photons per second used was in principle high enough to be able to reach the SQL at frequencies close to the mechanical resonances. The mechanical oscillators in Refs.~\cite{Teu09,Kipp2009} were nano-wires of aluminum and SiN, respectively, with effective masses of a few picograms and resonance frequencies in the MHz regime.

Here, we report on an opto-mechanical experiment with a readout noise below the SQL, similar to Refs.~\cite{Teu09,Kipp2009}. In contrast to previous works, our setup was composed of an interferometer without cavities and a  commercially available mechanical probe -- a SiN$_{\rm{x}}$ membrane with an effective mass of about 100\,ng and a resonance frequency of about 100\,kHz. The membrane was used as a partially transmitting mirror inside a free-space Michelson-Sagnac interferometer \cite{Kazuhiro10,Daniel11}. The experiment was performed at room temperature and achieved a total readout noise of $3 \times 10^{-16}\, \rm{m}/\!\sqrt{\rm{Hz}}$. The linear spectral density value was a factor of more than 6 (the power spectral density $16\,\rm{dB}$) below the peak value of the SQL.

The quantum readout noise of an interferometer is given by photon counting noise, also called shot-noise (sn). A rather useful measure is $\sqrt{S_{x\rm{,sn}}}$, the linear spectral density of the shot-noise calibrated to an apparent displacement $x$, given in $\rm{m}/\!\sqrt{\rm{Hz}}$. Since the shot-noise has a white spectrum, and our interferometer has a white signal transfer function, the normalized spectral density is frequency independent and reads \cite{Kazuhiro10}
\begin{equation}
	\sqrt{S_{x,\rm{sn}}}=\sqrt{\frac{\hbar c \lambda}{16\pi r^2 P}}\, ,
	\label{SN}
\end{equation}
with $\hbar$ the reduced Planck constant, $c$ the speed of light, $\lambda$ the laser wavelength, $r^2$ the membrane's power reflectivity, and $P$ the light power inside the interferometer. Note that the membrane is serving as a common end-mirror thereby doubling the displacement signal compared to a single movable end-mirror in an ordinary Michelson interferometer. 

The quantum back-action noise of an interferometer is given by the quantum radiation-pressure noise (rpn) on the mechanical probe. The susceptibility of a mechanical oscillator reads 
\begin{equation}
	\chi (\Omega) =  \frac{x(\Omega)}{F(\Omega)} =  \frac{-1}{m(\Omega^2 - \rmi \Omega\,\Omega_{\rm{m}}/Q -\Omega_{\rm{m}}^2)}  \text{\quad ,}
	\label{chi}
\end{equation}
with $x(\Omega)$ the spectrum of probe's displacement, $F(\Omega)$ the spectrum of the external force, $m$ the effective mass of the mechanical probe, $Q$ the mechanical quality factor, and $\Omega_{\rm{m}} = 2\pi f_{\rm{m}}$ the oscillator eigenfrequency. Far above $\Omega_{\rm{m}}$, the susceptibility is proportional to $\Omega^{-2}$. For frequencies far below the resonance, $\chi$ is constant, whereas its absolute value on resonance is given by $|\chi_{\rm{peak}}|=Q/(m \Omega_{\rm{m}}^2)$. Due to the frequency dependence in Eq.~(\ref{chi}), the signal normalized spectral density of the back-action noise is not white. For the membrane in our Michelson-Sagnac interferometer it is given by \cite{Kazuhiro10}
\begin{equation}
	\sqrt{S_{x,\rm{rpn}}(\Omega)}=|\chi (\Omega)| \, \sqrt{\frac{16\pi\hbar r^2 P}{c\lambda}}\, .
	\label{RPN}
\end{equation}

The quantum noise spectral densities given in Eqs.~(\ref{SN}) and (\ref{RPN}) both set limitations on the precision of a force measurement: while the shot-noise originates from the phase quadrature uncertainty of the light inside the interferometer, the radiation-pressure noise originates from its amplitude quadrature uncertainty \cite{Cav81}. Assuming that the two quadratures are uncorrelated with each other, which is the case for coherent states, their noise variances simply add up, $S_{x,tot}=S_{x, sn}+S_{x, rpn}$. For every frequency $\Omega/2\pi$ there exists an optimum laser power $P_{\rm SQL} = c\lambda/(16|\chi(\Omega)|\pi r^2)$ at which the two spectral densities have the same values. In this case the uncorrelated sum is minimal defining the standard quantum limit with a displacement normalized spectral density of
\begin{equation}
\sqrt{S_{x,\rm{SQL}}(\Omega)} = \sqrt{2\hbar \,|\chi(\Omega)| \,}\, .
	\label{SQLx}
\end{equation}
Using Eq.~(\ref{chi}) the SQL force spectral density can be written as
\begin{equation}
\sqrt{S_{F,\rm{SQL}}(\Omega)}=\sqrt{\frac{2\hbar}{|\chi(\Omega)|}\,}\, .
	\label{SQLF}
\end{equation}
The spectral densities given above need to be reached {(at least at one observation frequency)} in order to be able to claim that the SQL has been reached.  
For instance, at the mechanical resonance frequency the total noise needs to be as low as the following SQL (peak) value 
\begin{equation}\label{SQLx_def}
\sqrt{S_{x,\rm{SQL}}(\Omega_{\rm{m}})} = \sqrt{ \frac{2\hbar Q}{m\Omega_{\rm{m}}^2}\,} .
\end{equation}

The mechanical oscillator used in our work was a $1.5\,\rm{mm}$ sized SiN$_{\rm{x}}$ membrane for x-ray spectroscopy \cite{norcada}. These high tensile-stress thin films deposited on a Si frame have previously been used as mechanical oscillators in opto-mechanical cavity setups \cite{Harris08a,Harris08b}.  
Their low effective mass combined with their rather high tensile-stress results in high fundamental resonance frequencies around $100\,\rm{kHz}$. For stoichiometric $\rm{Si}_3\rm{N}_4$ membranes, mechanical resonance frequencies are even higher and lie in the megahertz region \cite{WRPK09}. High tensile-stress SiN membranes are interesting mechanical oscillators since their large surface area makes them a truly macroscopic object. Their surface quality is quite high, i.e.~high enough to be able to place a membrane in the middle of an optical cavity with finesse 15,000 \cite{Harris08b} and also high enough to use the membrane in a high-precision interferometer as demonstrated here. 

In order to determine the parameters of our membrane, we measured its power reflectivity to $30.4\,\%$ at 1064\,nm. The reflectivity is directly linked to the membrane's index of refraction of $n\!=\!2.2+i\,1.5\times\!10^{-4}$ at 1064\,nm \cite{Harris08b} and  to its etalon effect, which we used to determine the membrane's thickness to 66\,nm. This yields an effective mass for the fundamental oscillation mode of $m = 0.25 \,\rho\, V \approx 115\,$ng, with mass density $\!\rho = 3.1 \,\rm{g}/\rm{cm}^3$ \cite{norcada2} and volume $V=(1.5\,\rm{mm})^2\!\times 66\,\rm{nm}$. To avoid residual gas damping of the membrane oscillation, our interferometer was set up in a vacuum chamber at a pressure of $10^{-6}\,\rm{mbar}$. The membrane's quality factor $Q$ was measured to $\approx10^6$. Residual gas damping was verified to reduce the measured $Q$ of the membrane by less than $0.3\,\%$.

\begin{figure}
	\includegraphics[width=8.6cm]{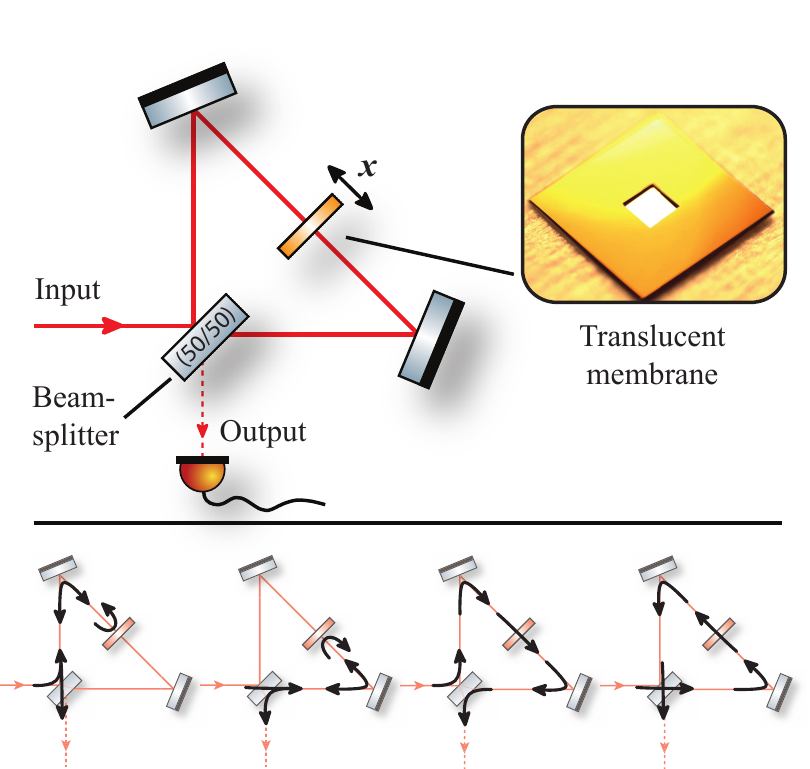}
	\caption{Schematic of the Michelson-Sagnac interferometer. The SiN$_{\rm x}$ membrane is aligned such that transmitted and reflected light beams propagate along the same optical axis. The lower row of plots show the four possible light paths, which interfere at the interferometer output. The inset shows a photograph of the membrane having an area of (1.5\,mm)$^2$.}
	\label{fig:exp}
\end{figure}
Figure~\ref{fig:exp} shows the Michelson-Sagnac topology of our free-space laser interferometer. It was invented to deal with the rather high light transmission of the membrane. The reflected as well as the transmitted light from the membrane is kept inside the same combined Michelson-Sagnac interferometer mode. First, the input laser beam is split up by a balanced beam splitter into two counter-propagating paths. Both paths are folded by highly reflective mirrors such that a shared focus lies at the membrane's position. The light fields reflected off the membrane interfere at the balanced beam splitter, forming the `Michelson mode' of our interferometer. The light beams transmitted through the membrane propagate along the same optical axis. They also overlap at the beam splitter and form the `Sagnac mode'. For a perfectly balanced beamsplitter, the fields of a Sagnac mode interfere at the beamsplitter in such a way that all the light goes back to the laser source. Light leaking out the other interferometer port then corresponds to the signal port of the Michelson mode. Its photo-electric detection provides information about the membrane's position. By placing the membrane such that the two Michelson arms have identical lengths, the signal port can be made completely dark limited only by the interference contrast. For a perfect dark port, the interferometer signal port shows a perfect rejection of laser intensity noise. In the present setup, however, we placed the membrane such that a small amount of light power leaked out the Michelson signal port in order to be able to detect the interferometer signal with a single photo diode. Note that in contrast to previous setups with membranes \cite{Harris08b,Marquardt} our present setup did not involve optical cavities; however, the Michelson-Sagnac topology in principle allows for the implementation of a power-recycling as well as a signal-recycling resonator for further sensitivity improvements \cite{Dre83,Mee88,Kazuhiro10}.

As light source we used a \emph{Mephisto} Nd:YAG laser that provided up to 2\,W light power at $\lambda \!=\! 1064$\,nm. 
The laser source was equipped with a direct feedback to the pump diode's current driver to suppress its relaxation oscillation (a so-called `noise eater'). For a further reduction of laser intensity noise, 10\,mW were picked off before the mode cleaner and detected on a photo diode. The signal derived was processed by a PID-controller and fed back to an acousto-optic modulator placed into the optical path before the mode cleaner. The servo's control bandwidth was about $1\,\rm{MHz}$ yielding an additional noise suppression of 12\,dB in the measurement band around 100\,kHz. As a result, technical laser noise did not influence the membrane displacement measurements reported here. 
About 53\,mW of the stabilized light did eventually circulate inside the interferometer. The photo-electric detection was performed with a single InGaAs photo diode of $500\,\rm{\mu m}$ diameter at the interferometer's signal port. The membrane position was piezo controlled such that about 0.55\,mW of light was detected. This light served as a local oscillator for the sideband fields produced by the membrane's oscillation appearing as an amplitude modulation of the local oscillator beam. 
\begin{figure}
	\includegraphics[width=8.4cm]{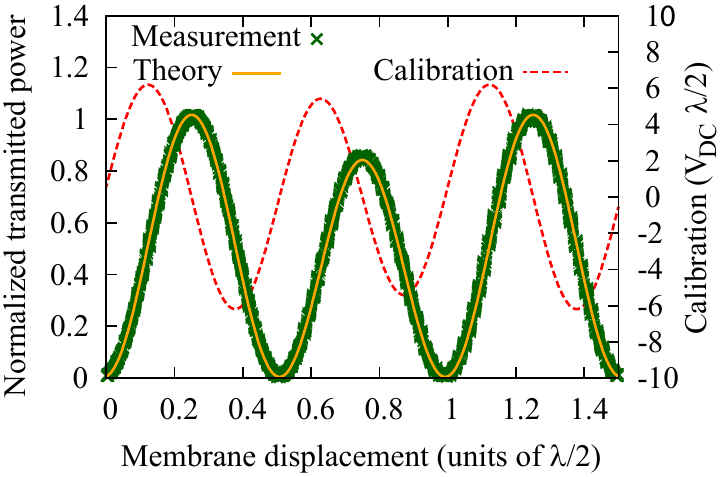}
	\caption{Comparison of the theoretical fringe pattern of a Michelson-Sagnac interferometer with measured output data. The derivative (dashed red line) of the transmitted power serves as calibration of the interferometer output in \fref{fig:spectrum}.}
	\label{fig:fringes}
\end{figure}

In the following paragraph we derive the Michelson-Sagnac interferometer's input-output relation, which is then used to model the fringe pattern at the signal port, and calibrate the interferometer's noise spectral density in $\rm{m}/\!\sqrt{\rm{Hz}}$. 
Let $\alpha_{\rm{in}}$ and $\alpha_{\rm{out}}$ be the complex amplitudes of the interferometer input and output fields, respectively, and $\rbs$ ($\rmem$) and $\tbs$ ($\tmem$) the amplitude reflectivity and transmissivity of the interferometer beam splitter (of the membrane), then the input-normalized output field at the signal port is given by
\begin{equation}
\begin{aligned}
\frac{\alpha_{\rm{out}}}{\alpha_{\rm{in}}}
	&=\rbs^2\tmem-\tbs^2\tmem\\
	&\quad+\rbs\tbs\rmem\exp{(i2\phi)}\\
	&\quad+\tbs\rbs\rmem\exp{(-i2\phi)}\\
	&=\left(\rbs^2-\tbs^2\right)\tmem+\rbs\tbs\rmem2\cos{(2\phi)}
	\label{Afringe}
\end{aligned}
\end{equation}
depending on the {membrane's detuning $\phi=2\pi \,x / \lambda$.}

The input-normalized light power at the photo diode is then given by
\begin{equation}
	\begin{aligned}
	\frac{P_{\rm{out}}}{P_{\rm{in}}}=\abs{\frac{A_{\rm{out}}}{A_{\rm{in}}}}^2
	=c_{\rm 1}+c_{\rm 2}\cos{(4\phi)}+c_{\rm 3}\cos{(2\phi)}\, ,
	\label{Pfringe}
         \end{aligned}
\end{equation}
with constants
\begin{equation}
	\begin{aligned}
	c_{\rm 1} &=\left(\rbs^2-\tbs^2\right)^2\tmem^2+2\rbs^2\tbs^2\rmem^2 \, , \\
	c_{\rm 2} &=2\rbs^2\tbs^2\rmem^2 \, , \\
	c_{\rm 3} &=4\left(\rbs^2-\tbs^2\right)\tmem\rbs\tbs\rmem \, .
	\label{Pfringeconstants}
	\end{aligned}
\end{equation}
The derivative 
\begin{equation}
\begin{aligned}
	\frac{\partial P_{\rm{out}}}{\partial x}
	&=-\frac{2c_{\rm 2} \pi P_{\rm{in}}}{\lambda}\sin{(4\phi)}-\frac{c_{\rm 3} \pi P_{\rm{in}}}{\lambda}\sin{(2\phi)}
\end{aligned}
\label{derivative}
\end{equation}
of this fringe equation provides the calibration of the interferometer output in photo-electric voltage per unit length.

Note that although the described interferometer can be thought of  as a composition of a Michelson and a Sagnac mode, their interference provides a distinct fringe pattern. A pure Sagnac interferometer can be dark at the output only if $t_{\rm{bs}}\!=\!r_{\rm{bs}}$, however, a Michelson-Sagnac interferometer can have a dark output port even if $t_{\rm{bs}}\neq r_{\rm{bs}}$. In the latter case the Michelson mode needs to be detuned from its own dark fringe to provide an overall dark fringe\cite{Daniel11}. The required differential phase of the Michelson mode is given by
\begin{equation}
	2 \phi^{\rm{dark}}=\arccos{\frac{\left(r^2_{\rm{bs}}-t^2_{\rm{bs}}\right)t_{\rm{m}}}{2r_{\rm{bs}}t_{\rm{bs}}r_{\rm{m}}}}\text{\quad .}
	\label{darkfringeoffset}
\end{equation}
Obviously, Eq.~(\ref{darkfringeoffset}) is {defined} only if the unbalancing of the beam splitter is not too strong. In our experiment the beam splitter was rather well balanced ($r_{\rm{bs}}/t_{\rm{bs}}=0.486/0.514$) and $\phi^{\rm{dark}}$ close to $\pi/4$.

Figure~\ref{fig:fringes} shows a measured fringe pattern of the Michelson-Sagnac interferometer (bold, green line). For the calibration of the x-axis we fitted a 2nd order polynomial function to the response function of the piezo that drove the membrane along the optical axis. The different heights of neighboring maxima is due to the slight unbalancing of the interferometer beam splitter. For comparison the theoretical fringe pattern according to \eref{Pfringe} is also given. The dashed curve in \fref{fig:fringes} shows its derivative according to \eref{derivative}. It was used for the absolute calibration of the interferometer's sensitivity. For a fixed input power, the fringe pattern and the DC output power directly provide the (DC) membrane position, i.e. the operating point of the interferometer, and the corresponding fringe derivative. The measured power spectrum of the high-pass filtered photo current was first divided by the independently measured AC-gain of the photo diode electronics and by the resolution bandwidth of the measurement in order to obtain the power spectral density of the photo-electric voltage in $\rm{V}^2\!/{\rm{Hz}}$. The result was then converted into a linear spectral density by taking the square root. Finally the calibration from Fig.~\ref{fig:fringes} was used to convert to a linear displacement noise spectral density.

\begin{figure}
	\includegraphics[width=8.6cm]{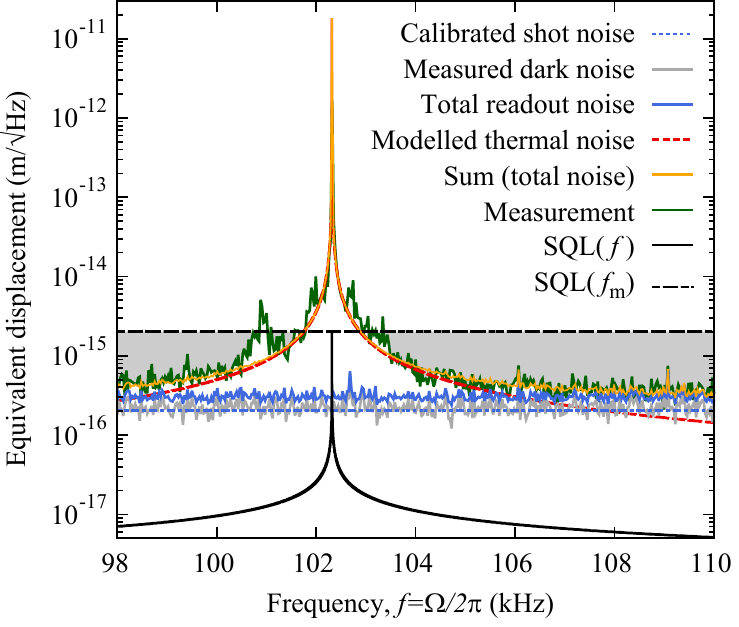}
	\caption{Measured (green) and modeled (yellow) noise spectrum of the Michelson-Sagnac laser interferometer containing a SiN$_{\rm{x}}$ membrane at room temperature. The total readout noise (imprecision) has a level of slightly above $3\times10^{-16}\,$m/$\sqrt{\rm{Hz}}$ and is given by the sum of electronic dark noise and shot-noise. It lies well below the peak value of the SQL of $2\times10^{-15}\,$m/$\sqrt{\rm{Hz}}$ at the membrane's resonance frequency $f_{\rm{m}}$. {For our modeling we assumed all noise sources being uncorrelated}. 
	}
	\label{fig:spectrum}
\end{figure}

Figure~\ref{fig:spectrum} shows the measured and modeled linear noise spectral densities of the membrane's displacement inside the Michelson-Sagnac interferometer. The spectral densities include technical and quantum noise of the interferometer's readout (shot-noise) and the room temperature thermal noise of the membrane. The shot-noise has a spectral density of $2\!\times\!10^{-16}\,\rm{m}/\!\sqrt{\rm{Hz}}$ and was derived from a noise measurement of the 0.55\,mW output field using the interferometer calibration described above. It was found in accordance to Eq.~(\ref{SN}) for an intra-interferometer light power of 53\,mW. The thermal noise was calculated following Ref.~\cite{Kazuhiro10}. Its absolute values agreed with the independent calibration of the interferometer's sensitivity described above.  
The spectral density of the membrane's SQL was calculated according to Eq.~(\ref{SQLx}). The peak value of the SQL is highlighted as a dashed horizontal line at a level of $2\times10^{-15}\,\rm{m}/\!\sqrt{\rm{Hz}}$. At off-resonant frequencies the measured noise spectral density corresponds to the readout noise (imprecision) being more than a factor $\sqrt{40}$ below the peak of the SQL. The broad peaks around the membrane resonance were due to beat signals with the mechanical resonances of the membrane mounting. 

To conclude, the Michelson-Sagnac interferometer of the present work shows a readout noise power spectral density (in $\rm{m}^2/\rm{Hz}$) being a factor of more than 40 (16\,dB) below the peak value of a SiN$_{\rm{x}}$ membrane's SQL. This factor is considerably higher than that achieved in previous works with nano-wires, in which factors of about 1.25 (1\,dB) \cite{Teu09} and 2 (3\,dB) \cite{Kipp2009,Kipp2010} were achieved, respectively. 
{The absolute calibration of the interferometer's sensitivity was based on the measurements of the absolute light power and its noise at the interferometer's signal port. It was found in very good agreement with the modeled thermal excitation of our room temperature membrane.} 
We emphasize that, just like all previous works, our opto-mechanical arrangement did not reach its SQL of a force measurement, leaving this an all-time goal. 
In order to reach the SQL at the membrane's resonance frequency we don't need to further reduce the absolute value of the readout noise, but rather the relative contribution of detector dark noise has to be lowered. But first of all, the membrane needs to be cooled to temperatures below $\hbar\Omega_{\rm{m}}/2k_{\rm{B}} \approx 2.5 \,\mu{\rm K}$, with $k_{\rm{B}}$ the Boltzmann constant. This rather low temperature is related to the rather low resonance frequency of the membrane. We note that the regime of quantum radiation pressure coupling should be possible at significantly higher temperatures around 1\,K \cite{Kazuhiro10}. 
In principle our interferometer can be enhanced by adding a mirror to the output port. 
In this configuration, the interferometer might enable the demonstration of dissipative cooling \cite{Xuereb11}. 
The same configuration could also be used to establish a signal recycling cavity \cite{Mee88,Kazuhiro10}. In this case the quantum readout noise is reduced without increasing the laser power at the membrane. However, the interferometer's broad-band signal-bandwidth would be reduced down to the signal-recycling cavity bandwidth. A lowered readout noise in our setup would certainly be valuable in order to enable the observation of the membrane's thermal noise spectrum below its resonance frequency. This might allow the identification of the underlaying loss channel in this kind of high-Q oscillator giving insight in solid state physics.

This work has been supported by the International Max Planck Research School for Gravitational Wave Astronomy and by the Centre for Quantum
Engineering and Space-Time Research, QUEST. We thank A. Xuereb, H. Miao and A. R{\"u}diger for helpful comments on our manuscript.


\begin{thebibliography}{99}
	\bibitem{Abramovici92} A.~Abramovici ~\textit{et al.}, Science \textbf{256}, 325--333 (1992). 
	\bibitem{Heisenberg} W.~Heisenberg, Zeitschrift f{\"u}r Physik 43 (3-4): 172--198 (1927).
	\bibitem{Bra1968} V.~B.~Braginsky, JETP {\bf 26}, 831 (1968).
	\bibitem{VGFBTGVZA07} D.~Vitali, S.~Gigan, A.~Ferreira, H.~R.~B{\"o}hm, P.~Tombesi, A.~Guerreiro, V.~Vedral, A.~Zeilinger, and M.~Aspelmeyer,  Phys. Rev. Lett. {\bf 98}, 030405 (2007). 
	\bibitem{Helge08} H.~M{\"u}ller-Ebhardt, H.~Rehbein, R.~Schnabel, K.~Danzmann, and Y.~Chen, Phys. Rev. Lett. \textbf{100}, 013601 (2008).
	\bibitem{Braginsky95} V.~B. Braginsky, F.~Y. Khalili, and K.~S. Thorne, \textit{Quantum measurement} (Cambridge University Press, Cambridge, England, 1995).
	\bibitem{KLMTV01} H.~J.~Kimble, Y.~Levin, A.~B.~Matsko, K.~S.~Thorne, and S.~P.~Vyatchanin, Phys. Rev. D {\bf 65}, 022002 (2001).
	\bibitem{Teu09} J.~D.~Teufel, T.~Donner, M.~A.~Castellanos-Beltran, J.~W.~Harlow, and K.~W.~Lehnert, Nature Nanotechnology {\bf 4}, 820--823 (2009).
	\bibitem{Kipp2009} G.~Anetsberger, O.~Arcizet, Q.~P.~Unterreithmeier, R.~Rivi{e}re, A.~Schliesser, E.~M.~Weig, J.~P.~Kotthaus, and T.~J.~Kippenberg, Nature Physics {\bf 5}, 909--914 (2009).
	\bibitem{norcada} www.norcada.com.
	\bibitem{Harris08a} J. D. Thompson, B. M. Zwickl, A. M. Jayich, F. Marquardt, S. M. Girvin, and J. G. E. Harris, Nature 452, 72 (2008).
	\bibitem{Harris08b} A.~M.~Jayich, J.~C.~Sankey, B.~M.~Zwickl, C.~Yang, J.~D.~Thompson, S.~M.~Girvin, A.~A.~Clerk, F.~Marquardt, and J.~G.~E.~Harris, New Journal of Physics {\bf 10}, 095008 (2008).
	\bibitem{WRPK09} D.~J.~Wilson, C.~A.~Regal, S.~B.~Papp, and H.~J.~Kimble, Phys. Rev. Lett. {\bf 103}, 207204 (2009)
	\bibitem{norcada2} Personal communication with Norcada.
	\bibitem{Daniel11} D.~Friedrich, H.~Kaufer, T.~Westphal, K.~Yamamoto, A.~Sawadsky, F.~Ya.~Khalili, S.~Danilishin, S.~Go{\ss}ler, K.~Danzmann, R.~Schnabel, New Journal of Physics {\bf 13}, 093017 (2011).
	\bibitem{Marquardt} F.~Marquardt and S.~M.~Girvin, Physics {\bf 2}, 40 (2009).
	\bibitem{Mee88} B.~J.~Meers, Phys. Rev. D {\bf 38}, 2317--2326 (1988).
	\bibitem{Kazuhiro10} K.~Yamamoto, D.~Friedrich, T.~Westphal, S.~Go{\ss}ler, K.~Danzmann, K.~Somiya, S.~L.~Danilishin, and R.~Schna\-bel, Phys. Rev. A {\bf 81}, 033849 (2010).
	\bibitem{Dre83} R.~W.~P.~Drever \textit{et al.}, in Quantum Optics, Experimental Gravity, and Measurement Theory, edited by P. Meystre and M. O. Scully (Plenum, New York, 1983), p. 503.
	\bibitem{Cav81} C.~M.~Caves, Phys. Rev. D {\bf 23}, 1693-1708 (1981).
  \bibitem{Kipp2010} G.~Anetsberger, E.~Gavartin, O.~Arcizet, Q.~P.~Unterreithmeier, E.~M.~Weig, M.~L.~Gorodetsky, J.~P.~Kotthaus, and T.~J.~Kippenberg, Phys. Rev. A {\bf 82}, 061804 (2010).
	\bibitem{Xuereb11} A. Xuereb, R. Schnabel, and K. Hammerer, \emph{Dissipative Optomechanics in a Michelson--Sagnac Interferometer}, Phys. Rev. Lett. to be published, arXiv:1107.4908.
\end{thebibliography}
\end{document}